\documentstyle[multicol,aps,epsf]{revtex}
\begin{document}
\draft

\title{Unusual $T_c$ variation with hole concentration in
Bi$_2$Sr$_{2-x}$La$_x$CuO$_{6+\delta}$}

\author{Mu-Yong Choi and J. S. Kim}

\address{Department of Physics and Institute of Basic Sciences \\
 Sungkyunkwan University, Suwon 440-746, Korea} 

\maketitle

\begin{abstract}

We have investigated the $T_c$ variation with the hole concentration $p$ in
the La-doped Bi 2201 system, Bi$_2$Sr$_{2-x}$La$_x$CuO$_{6+\delta}$. 
It is found that the Bi 2201 system does not follow the systematics in $T_c$ and $p$
observed in other high-$T_c$ cuprate superconductors (HTSC's). 
The $T_c$ vs $p$ characteristics are quite similar to what observed in Zn-doped HTSC's.
An exceptionally large residual resistivity component in the inplane resistivity 
indicates that strong potential scatterers of charge carriers reside in
CuO$_2$ planes and are responsible for the unusual $T_c$ variation with $p$, as in the
Zn-doped systems. However, contrary to the Zn-doped HTSC's, the strong scatter in the Bi
2201 system is possibly a vacancy in the Cu site.
\end{abstract}

\draft

\begin{multicols}{2}
Many high-$T_c$ cuprate superconductors (HTSC's) display an approximately parabolic
dependence of $T_c$ upon the hole concentration $p$ with the maximum $T_c$ at $p
\approx$ 0.16.\cite{R1,R2}  ($p$ is defined as the hole concentration per Cu
atom in CuO$_2$ planes. ) This behavior was observed first in
La$_{2-x}$Sr$_x$CuO$_4$.\cite{R1} Then other HTSC's such as
YBa$_2$Cu$_3$O$_{7-y}$\cite{R2}, Bi$_2$Sr$_2$CaCu$_2$O$_{8+\delta}$\cite{R3},  
and TlSr$_2$CaCu$_2$O$_{7+\delta}$\cite{R1} were also found to show
approximately the same relation between $T_c$ and $p$ which scales only with
the  maximum $T_c$, $T_{c,~max}$. Though not studied for the full range of $p$,
several other HTSC's are also known to have $T_{c,~max}$ at $p \approx$
0.14 $\sim$ 0.15.\cite{R4,R5} Therefore one might expect that there possibly
exists a universal relation between $T_c$ and $p$ which all HTSC's
satisfy.\cite{R2}

Existence of a universal parabolic relation between $T_c$ and $p$ for all HTSC's, despite
the different combinations of constituent atoms, the presence of various charge-carrier
reservoir layers, and a variety of inter-plane coupling strengths, 
cannot be common but is believed to be related to a noble nature of high-temperature
superconductivity. It is therefore not strange that the recent observations in Zn-doped
HTSC's of departure from the universal relation have drawn particular
interest.\cite{R6,R7,R8,R9} Much attention has focused on the function of Zn. Within
a HTSC, Zn substitutes for Cu in the CuO$_2$ plane and behaves as a nonmagnetic impurity
without altering the carrier concentration. In this report, we show that a similar
non-universal $T_c$-$p$ relation holds also for the La-doped Bi 2201 system,
Bi$_2$Sr$_{2-x}$La$_x$CuO$_{6+\delta}$, which contains strong disorders in CuO$_2$ planes
differing from impurities.

We have obtained the hole concentration $p$ of the samples from the thermopower
($S$) measurements. The room-temperature thermopower $S$(290 K) of HTSC's was
found to be a universal function of $p$ over the whole range of doping,
\cite{R1,R3} which has since been used widely to determine the $p$ of HTSC's.
The superconducting-transition temperature $T_c$ was determined at half the
normal-state resistivity. The conventional solid-state reaction of
stoichiometric oxcides and carbonates was adopted in preparing polycrystalline
samples of Bi$_2$Sr$_{2-x}$La$_x$CuO$_{6 + \delta}$. The x-ray diffraction (XRD)
analysis shows all the samples to be single phase to the threshold of
detection. The oxygen content in the sample of $x$ = 0.1 could be varied by 
annealing the same sample in vacuum for 6 h at different temperatures
(400$^{o} $C, 500$^{o}$C, and then 600$^{o} $C). $S$ was measured by
employing the dc method described in Ref. 10. The resistivity $\rho$ was
measured through the conventional low-frequency ac four-probe method. 

\begin{figure}
\narrowtext
\centerline{\epsfxsize=3.2in
\epsffile{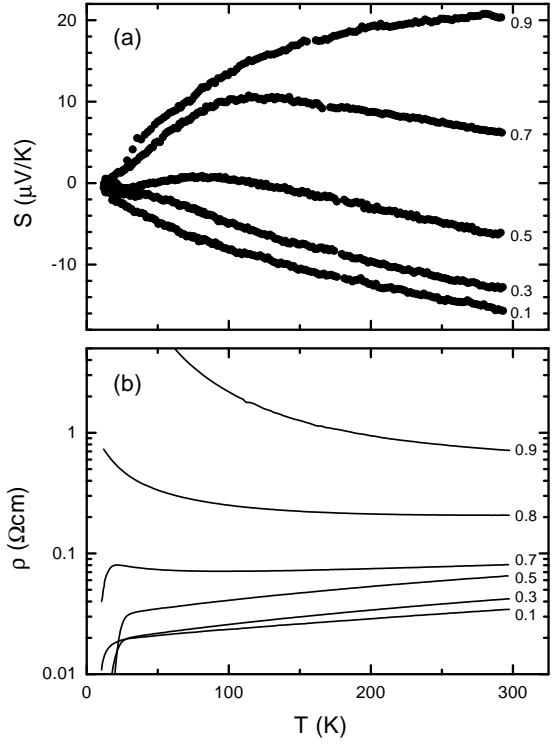}}
\vskip 0.1true cm
\caption{(a) The thermopower $S$ and (b) the resistivity $\rho$ of
Bi$_2$Sr$_{2-x}$La$_x$CuO$_{6+\delta}$ as functions of temperature. The
numbers next to the curves denote the La content $x$ in the materials.} 
\label{fig1}
\end{figure}

\begin{figure}
\narrowtext
\centerline{\epsfxsize=3.2in
\epsffile{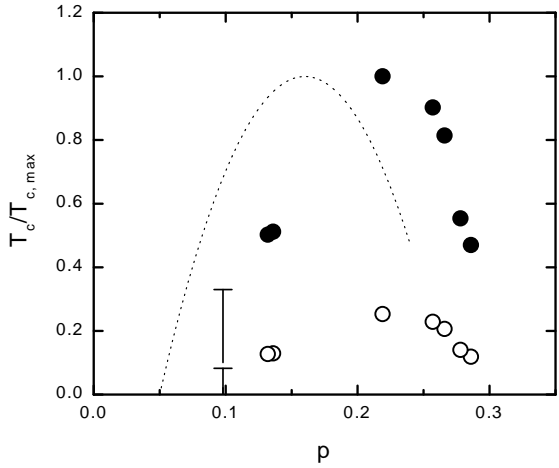}}
\vskip 0.1true cm
\caption{$T_c$ of Bi$_2$Sr$_{2-x}$La$_x$CuO$_{6+\delta}$, normalized to $T_{c,~max}$,
plotted as a function of the hole concentration $p$ determined from the $S$ data in Figure
1 and the $S$-$p$ relations in Ref. 1. $T_{c,~max}$ = 21.5 K for closed circles and 85 K
for open circles. The error bars show the upper limit of $T_c$ for the sample of $x$ =
0.8 with $p$ = 0.098. The dotted curve is a plot of the `universal' relation in Ref. 1.}  
\label{fig2} 
\end{figure}

Figure 1 shows the temperature dependences of $S$ and $\rho$ of
Bi$_2$Sr$_{2-x}$La$_x$CuO$_{6 + \delta}$ (BSLCO) with 0.1 $\le x \le$ 0.8.
The temperature and doping dependences of $S$ in Fig.\ \ref{fig1}(a) are typical of
HTSC's. $S$(290 K) increases with doping $x$ from -15.5 $\mu$V/K to 60 $\mu$V/K. 
Corresponding $p$ determined from the relations between $S$(290 K) and $p$ in
Ref. 3 varies from 0.286 to 0.073 with doping. The $\rho$ measurements in Fig.\
\ref{fig1}(b) displays that the $T_c$ of BSLCO has its maximum at $x \sim$ 0.5
or $p \sim$ 0.22. The appearance of $T_{c,~max}$ at $x \sim$ 0.5 agrees
with the previous measurements.\cite{R11} $T_c$/$T_{c,~max}$ against $p$ is
plotted in Figure 2. The $T_c$ (= 21.5 K) of $x$ = 0.5 is used as $T_{c,~max}$ for 
solid circles. The dotted curve is of the `universal' relation, 
$T_c$/$T_{c,~max}$ = 1- 82.6 ($p$ - 0.16)$^2$, in Ref. 1. 
The relation has not yet been fully tested in the
overdoped region of $p >$ 0.25. Figure\ \ref{fig2} clearly displays that BSLCO 
does not follow the systematics. Superconductivity in the underdoped
region is deeply suppressed and the $T_{c,~max}$ appears at an overdoped hole
concentration $p \sim$ 0.22 rather than 0.16 .  Besides, the $T_{c,~max}$ of
$\sim$ 21.5 K is also unusually low, which is only ${1 \over 4}$ the $T_c$ of
Tl$_2$Ba$_2$CuO$_{6 + \delta}$,  isostuctural of BSLCO.\cite{R12}
 Taking the maximum $T_c$ of Tl$_2$Ba$_2$CuO$_{6+\delta}$ as 
$T_{c,~max}$, BSLCO has much lower $T_c$/$T_{c,~max}$'s, as represented by 
open circles in Figure 2.

\begin{figure}
\narrowtext
\centerline{\epsfxsize=3.2in
\epsffile{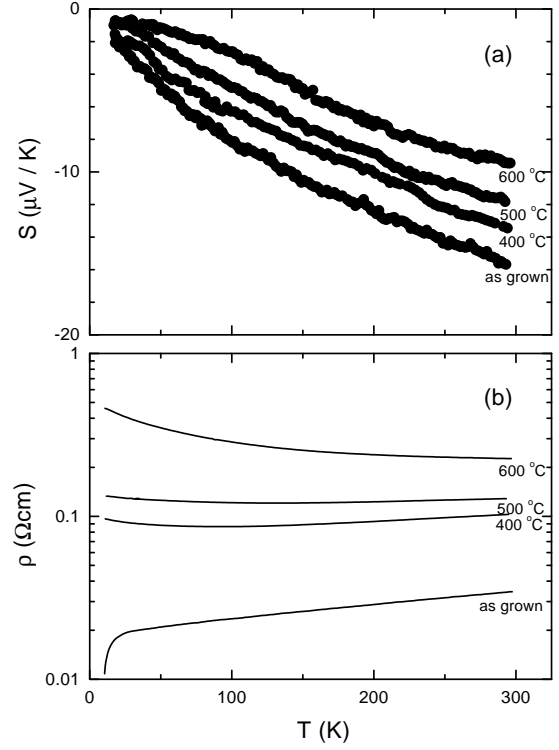}}
\vskip 0.1true cm
\caption{(a) $S$ and (b) $\rho$ of vacuum-annealed
Bi$_2$Sr$_{1.9}$La$_{0.1}$CuO$_{6+\delta}$ as functions of temperature. The
numbers next to the curves denote the annealing temperatures.}
\label{fig3}
\end{figure}

Unusual $T_c$ variation with $p$ is exposed more dramatically in
the vacuum-annealed sample of $x$ = 0.1 which superconducts at $T \le $ 10 K
without vacuum-annealed. Vacuum annealing reduces the content of oxygen atoms
interstitial between Bi-O planes and consequently $p$ in CuO$_2$
planes.\cite{R13,R14,R15} Fig\ \ref{fig3}(a) shows that successive vacuum
annealings at 400$^o$C, 500$^o$C, and then 600$^{o}$C enhance $S$ of
Bi$_2$Sr$_{1.9}$La$_{0.1}$CuO$_{6 + \delta}$ from -15.5 $\mu$V/K to -9.3
$\mu$V/K. The corresponding variation of $p$ is from 0.286 to 0.240. We expect
from the observed $T_c$-$p$ relation of BSLCO in Figure 2 that $T_c$ of the
sample of $x$ = 0.1 rises with annealing from 10 K to 20 K. The $\rho$
measurements in Figure\ \ref{fig3}(b), however, show that the superconductivity
observed in the as-grown sample disappears with annealing in vacuum. We observed
similar behaviors also in Bi$_2$Sr$_2$CuO$_{6+\delta}$ which had been prepared
from the nominal composition of Bi:Sr:Cu = 2:2:1.5. The semiconducting as-grown
sample of  Bi$_2$Sr$_2$CuO$_{6+\delta}$  having $p$ = 0.282 exhibited a
superconducting-transition onset at 11.5 K when vacuum-annealed at 400$^o$C.
And yet subsequent vacuum annealings at 500$^o$C and 600$^o$C put the sample
back in the semiconducting states. The $p$'s of the Bi$_2$Sr$_2$CuO$_{6+\delta}$
sample annealed at  400$^o$C, 500$^o$C, and 600$^o$C were 0.256, 0.250 and
0.216 respectively, all of which are located in the superconducting region of
Figure\ \ref{fig2}. 

The $T_c$ vs $p$ characteristics of as-grown samples represented by the open circles in
Figure\ \ref{fig2} resemble those of Zn-doped HTSC's in Ref. 6 and 7. It has been
suggested that  the primary effect of Zn impurities is to produce a large residual
resistivity as a nonmagnetic potential scatterer in the unitary limit and that the more
rapid depression of $T_c$ in the underdoped region is related to the large residual
resistivity reaching the universal two-dimensional resistance h/4e$^2$ $\approx$ 6.5
k$\Omega$/$\Box$ per CuO$_2$ plane at the edge of the underdoped superconducting region.
\cite{R8} Unlike most HTSC's, the Bi 2201 superconductor is found to have an exceptionally
large residual resistivity.\cite{R16,R17} The corresponding two-dimensional
residual resistance per CuO$_2$ plane ranges from 0.3 k$\Omega$/$\Box$
at an overdoped hole concentration to 10 k$\Omega$/$\Box$ at an underdoped concentration
with 50 \% uncertainties.\cite{R18} The large residual resistivity indicates that BSLCO
contains strong scatterers of charge carriers in the planes. The strong scatterer in
BSLCO is, however, not an impurity but most likely a vacancy in the Cu site, since any of
Bi, Sr, and La can hardly substitute for Cu and disorders in the noncopper sites have
little effect on superconducting properties but changing the hole concentration.
Nevertheless, a vacancy in the CuO$_2$ plane is expected to act as a nonmagnetic
potential scatterer, just like the Zn impurity in the planes.  
Vacuum annealing may cause extra vacancies in CuO$_2$ planes as well as expelling
intersititial oxygen atoms. Thus the same arguement in terms of disorder in the CuO$_2$
plane can be adopted for an explanation of the deeper suppression of $T_c$ in
vacuum-annealed samples.
  
Although the above discussion does not provide a full account for the origin of the
nonuniversal $T_c$ vs $p$ characteristics, it may be concluded that similarity between
the Bi 2201 HTSC with disorders differing from impurities and other HTSC's
with Zn impurities seem to strengthen the arguement that a strong
potential scattering in the planes and a large residual resistivity at an underdoped hole
concentration are closely related to the strong suppression of high-temperature 
superconductivity and the more rapid $T_c$ depression in the underdoped region.\\

 We wish to thank Y. Yun and I. Baek for their assistances with the XRD analysis.

\newpage

\end{multicols}

\begin{references}
\bibitem {R18}
Hole concentrations of the samples in Ref. 16 and 17 are estimated from the inplane
resistivity  at room temperature, $\rho_{ab}$(300 K), and the temperature dependence of
$\rho_{ab}$. A $\rho_{ab}$(300 K) $\approx$ 5 m$\Omega$cm and a semiconductor-like
temperature dependence at low temperatures of $\rho_{ab}$ usually appear in a sample with
an underdoped hole concentration.
\end{references}
\end{document}